\documentclass{iaus}
\usepackage{graphicx}

\title[How do methanol masers appear?] %% give here short title %%
{How do methanol masers manage to appear in the youngest star
vicinities and isolated molecular clumps?}

\author[Sobolev \etal]   %% give here short author list %%
{A.M. Sobolev$^1$, D.M. Cragg$^2$, S.P. Ellingsen$^3$, M.J.
Gaylard$^4$, S.~Goedhart$^4$, C. Henkel$^5$, M.S. Kirsanova$^6$,
A.B. Ostrovskii$^1$, N.V.~Pankratova$^1$, O.V. Shelemei$^1$, D.J.
van der Walt$^7$, T.S.~Vasyunina$^8$, M.A. Voronkov$^9$ }

\affiliation{$^1$Ural State University, Ekaterinburg, 620083,
Russia, email: Andrej.Sobolev@usu.ru\\[\affilskip]
$^2$Monash University, Clayton, Australia, email:
Dinah.Cragg@monash.edu.au\\[\affilskip]
$^3$University of Tasmania, Hobart, Australia, email:
Simon.Ellingsen@utas.edu.au\\[\affilskip]
$^4$Hartebeesthoek Radio Astronomy Observatory, South Africa,
email:
sharmila@hartrao.ac.za\\[\affilskip]
$^5$MPIfR, Bonn, Germany, email:
p220hen@mpifr-bonn.mpg.de\\[\affilskip]
$^6$Institute for Astronomy, Moscow, Russia, email:
kirsanova@inasan.ru\\[\affilskip]
$^7$North-West University, South Africa, email:
johan.vanderwalt@nwu.ac.za\\[\affilskip]
$^8$MPIA, Heidelberg, Germany, email:
Vasyunin@mpia-hd.mpg.de\\[\affilskip]
$^9$ATNF CSIRO, Sydney, Australia, email: Maxim.Voronkov@csiro.au
}

\pubyear{2007}
\volume{242}  %% insert here IAU Symposium No.
\date{?? and in revised form ??}
\jname{Astrophysical Masers and their Environments}
\editors{J. M. Chapman \& W. A. Baan, eds.}
\begin{document}

\maketitle

\begin{abstract}
General characteristics of methanol (CH$_3$OH) maser emission are
summarized. It is shown that methanol maser sources are
concentrated in the spiral arms. Most of the methanol maser
sources from the Perseus arm are associated with embedded stellar
clusters and a considerable portion is situated close to compact
H{\sc ii} regions. Almost 1/3 of the Perseus Arm sources lie at
the edges of optically identified H{\sc ii} regions which means
that massive star formation in the Perseus Arm is to a great
extent triggered by local phenomena. A multiline analysis of the
methanol masers allows us to determine the physical parameters in
the regions of maser formation. Maser modelling shows that
class~II methanol masers can be pumped by the radiation of the
warm dust as well as by free-free emission of a hypercompact
region (hcH{\sc ii}) with a turnover frequency exceeding 100~GHz.
Methanol masers of both classes can reside in the vicinity of
hcH{\sc ii}s. Modelling shows that periodic changes of maser
fluxes can be reproduced by variations of the dust temperature by
a few percent which may be caused by variations in the brightness
of the central young stellar object reflecting the character of
the accretion process. Sensitive observations have shown that the
masers with low flux densities can still have considerable
amplification factors. The analysis of class~I maser surveys
allows us to identify four distinct regimes that differ by the
series of their brightest lines.

\keywords{masers; catalogs; surveys; stars: formation; ISM:
clouds, evolution, kinematics and dynamics, structure; Galaxy:
structure; radio lines: ISM}
\end{abstract}

\firstsection % if your document starts with a section,
\section{Introduction}\label{sec:int}

Methanol is a complex molecule~--- it consists of six atoms, and two
of them are heavy. It is surely the 1st prize winner by complexity
among the molecules producing cosmic masers. It is really
surprising that such a molecule gives birth to numerous and
extremely bright masers that are only matched by the 1.7~GHz
hydroxyl (OH) and 22~GHz water vapor (H$_2$O) lines.

This paper describes the methanol maser phenomenon on different
spatial scales emphasizing the physical conditions that
characterize methanol maser sites. Several aspects of this topic
are covered in papers of other participants of this conference. We
concentrate on results obtained by researchers from the Ural State
University mostly in cooperation with scientists from other
institutes.

\section{General characteristics}\label{sec:gen}

Methanol masers are traditionally divided into two classes
according to the sets of transitions in which they emit.
Historically, the first methanol masers (later attributed to
class~I) were discovered in Orion in the $J_2-J_1$~E series of
lines at about 25~GHz (\cite[Barrett \etal\ 1971]{bar71}). The
existence of the second class of methanol masers (eventually named
class~II) was justified in the paper by \cite{bat87} after the
discovery of strong  masers in the $2_0-3_{-1}$~E line at
12.1~GHz. The existence of an intermediate class of methanol
masers is predicted but these masers are not likely to attain
fluxes comparable to those of the strongest lines of the two major
classes (\cite[Voronkov \etal\ 2005]{vor05}). Methanol masers of
both classes often co-exist in the same star-forming region,
though there are bright and notorious examples of class-I-only
(OMC-1) and class-II-only (W3(OH)) sources.

Methanol maser lines can achieve extremely high brightness:
class~I maser lines \linebreak at~44~GHz are reported to be
brighter than $10^8$~K (\cite[Kogan \& Slysh 1998]{kog98}) and
class~II lines at~6.7~GHz achieve even higher brightness exceeding
$10^{12}$~K (\cite[Menten \etal\ 1992]{men92}). Spectra of the
bright methanol masers usually contain numerous components (see,
e.g., \cite[Caswell \etal\ 1995]{cas95}) spread in a range smaller
than 20~km/s around the systemic velocity of the respective
molecular core (\cite[Malyshev \& Sobolev 2003]{mal03}). Maser
components are usually narrow (about 0.5~km/s) and the narrowest
ever reported line ($<$0.1~km/s) belongs to a~class~I methanol
maser (\cite[Voronkov \etal\ 2006]{vor06}). Methanol masers
display time variability of various kinds including periodic,
chaotic and episodic flaring (\cite[Goedhart \etal\ 2004]{goe04}).

\section{Methanol masers on a Galactic scale}\label{sec:gal}

Studies of the distribution of methanol masers in the Galaxy can
be reduced to class~II maser sources because most of the class~I
maser sites are found close to class~II ones (see, e.g.,
\cite[Slysh \etal. 1994]{sly94} and \cite[Ellingsen 2005]{ell05}).
At present there exist three unified catalogues of class~II
methanol masers each containing information on more
than~500~objects (\cite[Malyshev \& Sobolev 2003]{mal03};
\cite[Xu, Zheng \& Jiang 2003]{xu03} and \cite[Pestalozzi \etal\
2005]{pest05}). The distinguishing feature of the \cite{mal03}
catalogue is that it contains data on several maser transitions
and cross-references to the data on molecular shock tracers. This
catalogue is updated on a monthly basis and is available at
http://astroserver.astro.usu.ru.

Most of the class~II maser sources were found toward positions of
OH masers and IRAS sources with specific spectral characteristics.
A~new way of targeting based on GLIMPSE data is suggested in
\cite{ell07}. Blind surveys, including a recent Arecibo survey by
\cite{pan07} led to the detection of many new maser sources in
selected areas. A~real breakthrough is expected after the
completion of the blind survey within the MMB (Methanol Multibeam)
project (see the paper by J.~Green \etal\ in this volume).

Information on the Galactic distribution of class~II methanol
masers was presented in several papers which established that
these masers mostly reside in the molecular ring of our Galaxy
(for one of the most recent articles, see \cite[Pestalozzi \etal\
2005]{pest05}). At present, class~II masers were found only in
high mass star-forming regions (\cite[Minier \etal\ 2003]{min03})
and it is natural to expect that these sources should be
concentrated in the spiral arms.

\begin{figure}
%\vspace{-10cm}
\includegraphics[width=5in,angle=0]{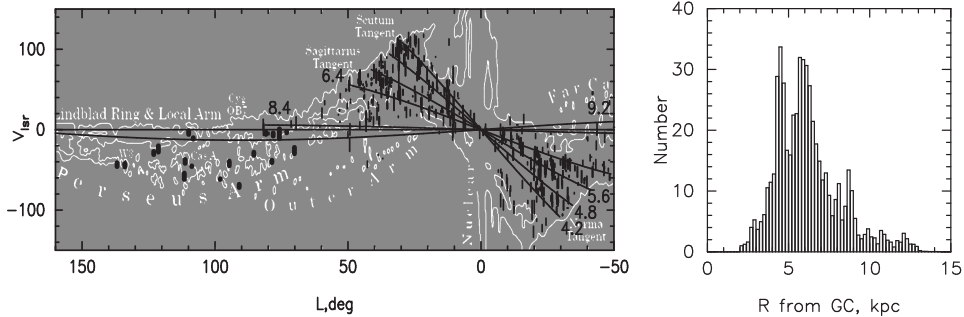}
\caption{Class~II methanol masers in the $V_{LSR}-l$ diagram
(lines of equal Galactocentric distances are shown, background: CO
data from \cite[Dame \etal\ 2001]{dam01}) and distribution of
the~maser sources as a function of Galactocentric
distance.}\label{fig:fg1}
\end{figure}

For the outer Galaxy, the concentration of methanol maser sources
to the Perseus spiral arm is quite certain: the velocity-Galactic
latitude ($l$) diagram shows that the masers mostly reside in
spiral arms  which are well outlined by CO emission (see
figure~\ref{fig:fg1}). Evidence for a concentration of the
methanol masers in the spiral arms is more difficult to prove in
the tangled regions of the inner Galaxy and was first provided in
\cite{sob05}. New maser sources were reported since then, mainly
in \cite{pan07} and \cite{ell07}. Figure~\ref{fig:fg1} shows a
$V_{LSR}-l$ diagram for the masers from the inner part of the
Galaxy along with lines of equal Galactocentric distance, $R$, and
the distribution of the maser sources as a function of $R$. The
updated \cite{mal03} catalogue and the \cite{bra93} rotation curve
were used. The concentration of masers to certain ranges of
Galactocentric distance is apparent. The values of $R$ correlate
with the nearest spiral arm sections of the inner Galaxy. It is,
however, not possible to conclude that all class~II masers reside
in the spiral arms. For the bulk of the sources there are only
kinematic distances which are not accurate enough. Moreover, most
of the distance estimates are based on velocities of 6.7~GHz maser
peaks which are significantly scattered with respect to the
systemic velocities of the clouds (\cite[Malyshev \& Sobolev
2003]{mal03} and \cite[van der Walt, Sobolev \& Butner
2007]{vdw07}). So a sophisticated analysis of the Galactic
methanol maser distribution requires a drastic improvement in the
precision of distance estimates.

\vspace{-3mm}
\section{Masers in the Perseus Arm}\label{sec:per}

\begin{figure}[t]
\begin{center}
\includegraphics[width=3.5in,angle=0]{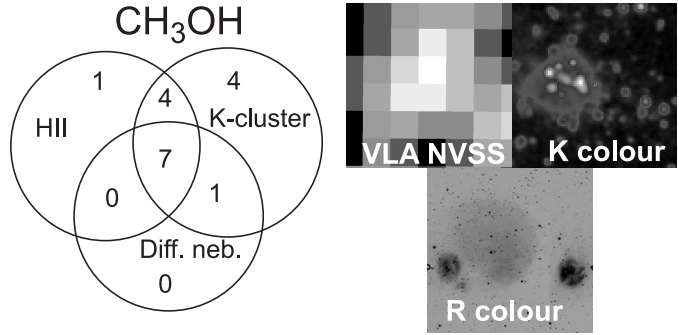}
\caption{Statistics of the association of class~II maser sources
from the Perseus Arm with optically visible H{\sc ii} regions,
embedded stellar clusters, and VLA NVSS sources.}\label{fig:fg2}
\end{center}
\end{figure}

In this section we will discuss the association of methanol masers
with the large scale constituents of star forming complexes in the
Perseus arm. We confine ourselves to the Perseus Arm because these
objects have relatively simple structure, most of them have no
confusing and strong back- or foreground and the objects are
relatively near. We realize that the Perseus Arm complexes are
somewhat different from those residing in the other parts of the
Galaxy but it is easier to start with the least complicated case.
28~sources were selected from the \cite{mal03} catalogue on the
basis of their position in the $V_{LSR}-l$ diagram. Using the
SkyView virtual observatory (http://skyview.gsfc.nasa.gov/) we
compiled an atlas of images of these sources displaying their
association with developed H{\sc ii} regions seen at optical
wavelengths (R~colour images), embedded star clusters revealed in
near infrared (K$_s$ colour images) and compact H{\sc ii} regions
detected in the NRAO VLA Sky Survey (NVSS). The numbers of maser
sources associated with these types of objects are shown in
figure~\ref{fig:fg2}. It is found that 57~\% of the sources are
associated with embedded clusters which means that these maser
sites can be affected by several young stellar objects (YSOs).
This is especially true for class~I masers which are found apart
from the YSOs. The fraction of cluster associations can be
substantially higher since K$_s$ colour images do not show very
deeply embedded clusters revealed at submm and mm wavelengths.
For~43~\% of the targets we find an association with VLA sources
which means that a~considerable fraction of maser sources is
associated with rather developed compact H{\sc ii} regions around
luminous YSOs. 29~\% of the sources appear to be associated with
optical H{\sc ii} regions. We conclude that a considerable
fraction of the methanol masers is triggered by the expansion of
evolved H{\sc ii} regions. Since the time duration of the methanol
masing phase is quite short (see, e.g., \cite[van der Walt \etal\
2005]{vdw05}) this means that contemporary massive star formation
in the Perseus Arm is to a great extent controlled by local
phenomena. A significant fraction of the sources ($\sim$40~\%,
mostly the distant ones) did not show a clear association with the
objects of our atlas.

\section{Class~II Methanol maser environments and pumping}\label{sec:classII}

The environment of class~II methanol masers was discussed in many
papers (see, e.g., \cite[Ellingsen 2006]{ell06}, \cite[Minier
\etal\ 2005]{min05} and \cite[Walsh \etal\ 2001]{wal01}) and talks
at this symposium (review by V. Fish, talks by J.~de Buizer, S.
Ellingsen, T. Hill, S. Longmore, L. Moscadelli and others). Here
we will instead emphasize two facts which are decisive for our
understanding of the maser pumping: it is established that almost
all of the maser sources are associated with mid-infrared (MIR)
emission (\cite[Walsh \etal\ 2001]{wal01}) and some of them are
associated with submm sources without cm - IR counterparts
(\cite[Hunter \etal\ 2006]{hun06}).

\subsection{Class~II methanol maser pumping regime}\label{sec:classIIreg}

Both observations and modelling of class~II masers show that the
6.7~GHz transition always manifests the highest brightness
temperature (\cite[Malyshev \& Sobolev 2003]{mal03}; \cite[Sobolev
\etal\ 1997]{sob97}). Observations have not shown significant
differences in the positions of the maser spots seen in different
transitions (see, e.g., \cite[Menten \etal\ 1992]{men92};
\cite[Sutton \etal\ 2001]{sut01}). So, there is basically only one
known pumping regime of class~II masers. However, ratios of the
brightnesses of different maser lines show considerable dependence
on the values of the physical parameters. The hypothesis that the
masers are pumped by MIR emission of the warm dust allows us to
explain observed brightness temperatures and to determine physical
parameters of several sources on the basis of multi-transitional
observations (e.g., \cite[Sutton \etal\ 2001]{sut01}, \cite[Cragg,
Sobolev \& Godfrey 2005]{cra05}). A model of the ``common''
class~II methanol maser source was constructed on the basis of
extensive surveys (e.g., \cite[Cragg \etal\ 2004]{cra04} and
\cite[Ellingsen \etal\ 2004]{ell04}). The results of such studies
are summarized in \cite{sob02} and \cite{cra05} and will not be
presented here.

\subsection{Class~II methanol maser variability}\label{sec:classIIvar}

The response of the maser fluxes to changes in the physical
parameters allows us to determine these parameters and their
variation in time via detailed modelling. For example,
interferometric studies of variability in the source G9.62+0.20
shows that different parts of the source are synchronized,
suggesting a radiative origin for the variability (\cite[Goedhart
\etal\ 2005]{goe05}). Monitoring observations have shown that this
source is almost strictly periodical (see the papers by Goedhart
\& Gaylard and Gaylard \& Goedhart in this volume). Periodicity is
revealed in the variations of G9.62+0.20 in three maser
transitions at 6.7, 12.2 and 107~GHz (van~der~Walt \etal\, in
preparation). A possible cause for the synchronized variability
might be a variation of the dust temperature in the region where
maser is formed. Model calculations show that corresponding
changes of fluxes can be reproduced by variations of the dust
temperature by a few per cent which may follow variations in the
brightness of the central YSO reflecting the character of the
accretion process.

\subsection{Class~II methanol maser pumping in hypercompact H{\sc ii} region environments}\label{sec:classIIhcHII}

\begin{figure}[b]
\begin{center}
%\vspace{-10cm}
\includegraphics[width=5in,angle=0]{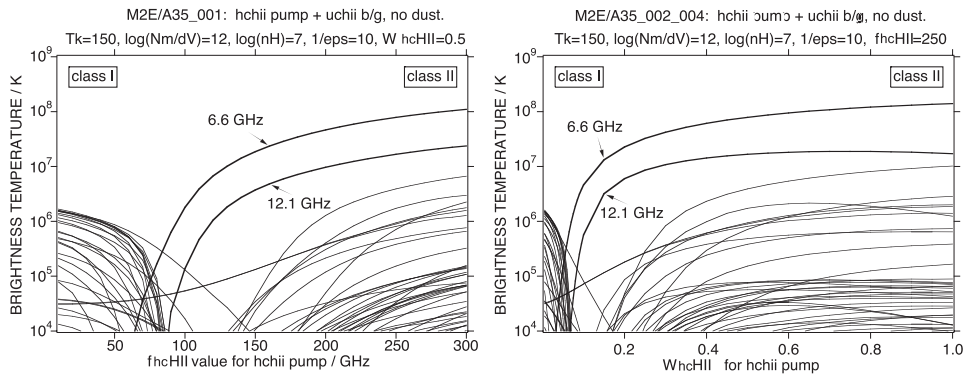}
\caption{Pumping methanol masers with free-free emission from a
hypercompact H{\sc ii} region. Dependence of maser brightnesses on
turnover frequency and dilution factor of the hcH{\sc
ii}.}\label{fig:fg3}
\end{center}
\end{figure}

Recent observations have shown that the masers which were
previously ``hanging in nowhere'' (e.g., the brightest 6.7~GHz
masers in NGC6334F, \cite[Ellingsen \etal\ 1996]{ell96}) are
associated with submm emission (\cite[Hunter \etal\ 2006]{hun06})
and have high chances to be formed in the vicinity of hypercompact
H{\sc ii} regions (hcH{\sc ii}s). These objects manifest strong
free-free emission which in some cases dominates the dust emission
up to frequencies as high as~300~GHz (\cite[Kurtz 2005]{kur05}).
We have undertaken model calculations in order to explore whether
methanol masers can be pumped by free-free emission with high
turnover frequency (i.e., high emission measure) without
considerable dust content. Historically, first calculations of
this kind were presented in the paper by (\cite[Slysh \etal\
2002]{sly02}) but these authors did not account for very powerful
torsional transitions which are crucial for pumping, at least in
terms of intensities. Our calculations are based on the most
recent transition rates that were used in the model calculations
presented in \cite{cra05}. The model parameters are close to those
presented by \cite{cra05} with the major difference that the warm
dust emission is substituted by the free-free emission from the
hcH{\sc ii}. We have chosen the following set of parameters: gas
temperature $T_k=150$~K, hydrogen number density
$n_H=10^7~$cm$^{-3}$, methanol specific column density
$N_m/dV=10^{12}~$cm$^{-3}$\,s, beaming factor $1/\epsilon=10$,
hcH{\sc ii} electron temperature $T_e=12\,000$~K, turnover
frequency $f_e=250$~GHz and hcH{\sc ii} dilution factor $W_{H{\sc
ii}}=0.5$. The calculations show that bright class~II masers are
indeed produced.

%\newpage
Figure~\ref{fig:fg3} presents results from two series of models.
The left panel reveals variations of brightness temperature when
modifying the turnover frequency. It is clearly seen that to pump
bright class~II masers free-free emission of hcH{\sc ii} should
have $f_e>100$~GHz, i.e. it should posess a very high emission
measure. The right panel demonstrates that the~pumping of class~II
methanol masers takes place quite close to the hcH{\sc ii} where
the~free-free emission is not greatly diluted ($W_{H{\sc
ii}}>0.1$). It is noteworthy that, when the~distance from the
hcH{\sc ii} increases ($W_{H{\sc ii}}\ll 0.1$), some dense, warm
and elongated clumps created by interaction of the hcH{\sc ii}
with the ambient medium can give birth to rather strong class~I
methanol masers. This is one of the possible explanations of the
fact that masers of different classes can reside close to each
other (see, e.g., \cite[Ellingsen 2005]{ell05}).

\subsection{Are weak masers really masing?}\label{sec:classIIweak}

\begin{figure}[b]
\begin{center}
%\vspace{-10cm}
\includegraphics[width=5in,angle=0]{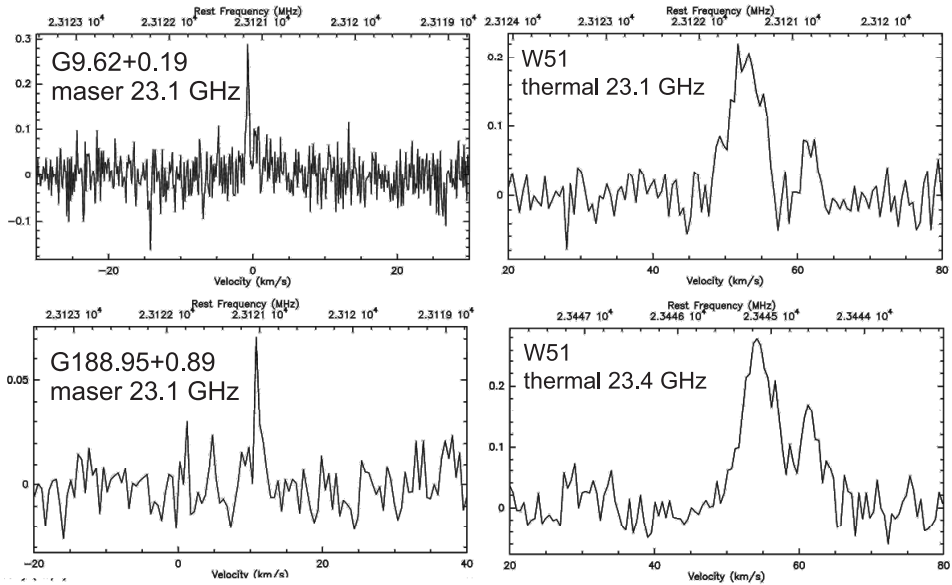}
\caption{Maser emission at 23.1~GHz and ``quasi-thermal'' emission
at 23.1 and 23.4~GHz.}\label{fig:fg4}
\end{center}
\end{figure}

It was shown that important constraints on the models can be
obtained through an evaluation of upper limits for the fluxes in
particular maser lines. Within this context it is very important
to know whether the weak lines are masers or whether a given
transition is thermalized. In order to find a solution to this
problem we conducted a very sensitive search for emission in the
$9_2-10_1\textrm{A}^+$ and $10_1-9_2\textrm{A}^-$ lines at 23.1
and 23.4~GHz using the~100-m telescope at Effelsberg. The
transitions at 23.1 and 23.4~GHz highly differ in their response
to factors producing departures from LTE: e.g., W3(OH) is strongly
masing at 23.1~GHz and shows absorption at 23.4~GHz
(\cite[Menten \etal\ 1986]{men86}).

In contrast to previous surveys at Parkes and Onsala the detection
rate in the maser line at~23.1~GHz was rather high and we detected
emission in 10 sources out of 20. The 23.1~GHz line in the~sources
W51, W75N, DR21(OH), Cep~A and G12.91--0.26 showed linewidths
comparable to those of corresponding ``quasi-thermal'' lines from
these objects. In these sources, the~23.4~GHz line shows very
similar intensities and line shapes (see the W51 spectra in
figure~\ref{fig:fg4}) which means that the transitions are
thermalized.

Toward the sources G9.62+0.19, G23.01+0.41 and G188.95+0.89 a
clear detection of the 23.1~GHz line was not accompanied by
23.4~GHz emission. The detected lines are considerably narrower
than ``quasi-thermal'' lines from these objects (see
figure~\ref{fig:fg4}). Line narrowing is a property of unsaturated
masers with considerable optical depth. So, we can state that we
detected three new weak but real masers and that the weakness of
masers does not necessarily mean that their amplification factors
are low.

\section{Class~I methanol maser environments and pumping}

The environment of class~I masers is highly turbulent. Lines of
the strong masers usually contain several spectral components and
their spots form clusters. General characteristics of the maser
images and spectra can be reproduced by modelling the turbulent
velocity field (\cite[Sobolev \etal\ 1998]{sww}). The existence of
the interferometric data on the images and time variability of the
\hbox{J$_2$ -- J$_1$E} maser line series at about 25~GHz in OMC-1
(\cite[Johnston \etal\ 1992]{joh92}) allow us to construct models
constraining the characteristics of the turbulence in this source
(\cite[Sobolev \etal\ 1998]{sww}; \cite[Sobolev \etal\
2003]{swo}).

Class~I methanol masers are less strong and widespread than
class~II ones. Most of them were found in the vicinity of class~II
maser sources (see, e.g., \cite[Slysh \etal\ 1994]{sly94} and
\cite[Ellingsen \etal\ 2005]{ell05}). There were extensive
searches (e.g., \cite[Slysh \etal\ 1994]{sly94} and \cite[Val'tts
\etal\ 2000]{val00}) for class~I sources but no unified catalogues
of these objects are published in the journals and no successful
blind surveys were performed yet. There are several lines of
evidence that class~I masers trace dense parts of molecular
outflows, probably locations where new stars might form (e.g.,
\cite[Voronkov \etal\ 2006]{vor06} and \cite[Sutton \etal\
2004]{sut04}) but generally the environment of class~I sources is
not well established yet.

Recently, sensitive and extensive surveys of the southern class~I
masers were conducted utilizing the unique capabilities of the
Australia Telescope Compact Array (see the paper by Voronkov
\etal\ in these proceedings). They allow us to draw some general
conclusions on the class~I maser pumping and identify four
distinct regimes differing by the series of the brightest (in
terms of brightness temperature) line.

The most frequently encountered maser regime shows the brightest
lines which belong to the  \hbox{J$_{-1}$ -- (J--1)$_0$E} series.
The 4$_{-1}-3_0$E and 5$_{-1}-4_0$E transitions at 36.1
and~84.5~GHz are weak masers under normal conditions of massive
star forming regions. Usually, the maser nature  of the lines in
this regime is difficult to prove observationally. Nevertheless,
there are cases where the line profiles contain narrow spikes and
the maser nature is proved interferometrically. The sources Sgr
B2, G30.8-0.1, and G1.6-0.025 can be considered as representatives
of this maser regime (\cite[Sobolev 1996]{sob96}).

In the second class~I maser regime  the lines of the \hbox{J$_0$
-- (J--1)$_1$A$^+$} series become prevalent. Numerous sources show
definitely maser lines arising in the \hbox{7$_0$ -- 6$_1$A$^+$}
and \hbox{8$_0$ -- 7$_1$A$^+$} transitions at 44.1 and 95.2~GHz,
respectively. Masers in the sources DR21W, NGC2264 and OMC-2
represent this regime (\cite{men91}). A preliminary theoretical
analysis of the pumping shows that the lines of the \hbox{J$_0$ --
(J--1)$_1$A$^+$} series become brightest in the~models with rather
high beaming ($>20$) and moderate column densities (\cite[Sobolev
\etal\ 2005]{sob05}).

The third maser regime is less widespread and is represented by
the sources where the lines of the \hbox{J$_2$ -- J$_1$E} series
at about 25~GHz are the brightest. OMC-1 is the prototypical
source (\cite[Johnston \etal\ 1992]{joh92})
and some additional bright sources were
found recently (Voronkov \etal\ these proceedings). Lines of this
series become brightest in the models with high specific column
densities ($lg(N_m/dV$~[cm$^{-3}\,\cdot\,$s])$\,\ge 12$) and
require relatively high temperatures ($T_k=75$\,--\,$100$~K) and
densities ($lg(n_H$~[cm$^{-3}])=5$\,--\,$7$).

The existence of the forth maser regime with the brightest lines
from the~\hbox{J$_{-2}$ -- (J--1)$_{-1}$E} series was previously
uncertain. Recent ATCA observations confirmed however, the
existence of such a regime in the sources W33-Met and
G343.12-0.06. Preliminary modelling shows that the \hbox{ 9$_{-2}$
-- 8$_{-1}$E} line at 9.9~GHz becomes brightest in models with
specific column densities which are as high as required for the
previously discussed regime. However, a smaller beaming angle and
either lower densities or higher temperatures ($ T_k>100$~K) are
also required.

\end{document}